\RequirePackage{etoolbox}
\let\originalshowhyphens\showhyphens

\documentclass[sigconf]{acmart}

\let\showhyphens\originalshowhyphens

\usepackage{graphicx}
\usepackage{tcolorbox}
\usepackage{multirow}
\usepackage{siunitx}
\usepackage{arydshln}
\usepackage[table]{xcolor}
\usepackage{booktabs}
\usepackage{url}

\tcbset{
    mysummarybox/.style={
        colback=blue!5!white,
        boxrule=0pt,
        colframe=blue!5!white,
        sharp corners,
        boxsep=5pt,
        left=0pt, right=0pt, top=0pt, bottom=0pt,
        fontupper=\normalsize
    }
}

\usepackage{enumitem}
\newlist{rquestions}{enumerate}{1}
\setlist[rquestions]{
    label=\textbf{RQ\arabic*:},
    labelwidth=*,
    font=\itshape,
    align=left
}

\copyrightyear{2026}
\acmYear{2026}
\acmConference[MSR '26]{23rd International Conference on Mining Software Repositories}{April 13--14, 2026}{Rio de Janeiro, Brazil}
\acmBooktitle{23rd International Conference on Mining Software Repositories (MSR '26), April 13--14, 2026, Rio de Janeiro, Brazil}

\begin{document}

\title{Fingerprinting AI Coding Agents on GitHub}

\author{Taher A. Ghaleb}
\orcid{0000-0001-9336-7298}
\affiliation{
  \department{Department of Computer Science} 
  \institution{Trent University}
  \city{Peterborough}
  \state{Ontario}
  \country{Canada}}
\email{taherghaleb@trentu.ca}

\begin{abstract}
AI coding agents are reshaping software development through both autonomous and human-mediated pull requests (PRs). When developers use AI agents to generate code under their own accounts, code authorship attribution becomes critical for repository governance, research validity, and understanding modern development practices. We present the first study on fingerprinting AI coding agents, analyzing 33,580 PRs from five major agents (OpenAI Codex, GitHub Copilot, Devin, Cursor, Claude Code) to identify behavioral signatures. With 41 features spanning commit messages, PR structure, and code characteristics, we achieve 97.2\% F1-score in multi-class agent identification. We uncover distinct fingerprints: Codex shows unique multiline commit patterns (67.5\% feature importance), and Claude Code exhibits distinctive code structure (27.2\% importance of conditional statements). These signatures reveal that AI coding tools produce detectable behavioral patterns, suggesting potential for identifying AI contributions in software repositories.
\end{abstract}



\maketitle

\section{Introduction}
AI coding agents such as GitHub Copilot, Devin, OpenAI Codex, Cursor, and Claude Code are now central to software development, assisting with code generation, testing, and autonomous pull request (PR) submission~\cite{li2025aidev}. While PRs created by autonomous agents are often explicitly labeled, developers also use AI agents to generate code but submit PRs under their own accounts.
This practice complicates repository governance, since projects may require disclosure of AI-generated code or restrict specific tools~\cite{github_tos}, yet enforcement depends on identifying AI contributions. It also threatens research validity, since datasets distinguishing ``\textit{human}'' and ``\textit{AI agent}'' PRs rely on submitter identity rather than actual authorship~\cite{li2025aidev}, risking mislabeled data and invalid conclusions. Further, it obscures our view of modern development practices~\cite{barke2023grounded}, impeding accurate measurement of AI's impact on code quality, productivity, and maintenance.

We hypothesize that AI coding agents leave detectable behavioral fingerprints in their PRs, even when mediated via human accounts, enabling code authorship attribution beyond submitter identity. To test this hypothesis, we conduct a study on fingerprinting AI coding agent, analyzing 33,580 PRs from five major AI agents in the AIDev dataset. Using 41 features spanning commit messages, PR structure, and code characteristics, we train a multi-class classifier that achieves a 97.2\% F1-score in identifying agents, revealing distinctive, agent-specific signatures in both text and code.

\medskip
\noindent\textbf{Contributions:}
This paper contributes:
(1) The first empirical study of AI coding agent fingerprinting, training supervised models to accurately detect undisclosed agent-generated PRs.
(2) Identification of distinctive behavioral signatures for five major AI coding agents, establishing fingerprints that enable reliable authorship attribution through feature- and model-level analysis.
(3) Implications for repository governance, AI agent design, and empirical software engineering, including risks to ``\textit{human}'' datasets such as AIDev.

\medskip
\noindent\textbf{Paper organization:}
Section~\ref{sec:empirical} presents our methodology and results for both research questions.
Section~\ref{sec:implications} discusses implications for maintainers, developers, and researchers.
Section~\ref{sec:threats} addresses validity threats.
Section~\ref{sec:related} reviews related work.
Section~\ref{sec:conclusion} concludes.

\medskip
\noindent\textbf{Replication:} Our data and scripts are available at~\cite{replication_package}.

\section{Empirical Analysis and Results}
\label{sec:empirical}

\subsection{Data Source}
We use the AIDev-pop dataset~\cite{li2025aidev} (Oct 28, 2025 update), containing PRs submitted by AI coding agents across GitHub repositories. The dataset contains 33,596 agentic PRs from five major AI coding agents: OpenAI Codex, GitHub Copilot, Devin, Cursor, and Claude Code. Each PR includes metadata (title, description, agent name, timestamps), commit information (messages, authors), and detailed code changes (file modifications, patches). 
After removing PRs with incomplete commit metadata, the final dataset has 33,580 PRs: OpenAI Codex (21,793, 64.9\%), GitHub Copilot (4,967, 14.8\%), Devin (4,822, 14.4\%), Cursor (1,540, 4.6\%), and Claude Code (458, 1.4\%). This natural class imbalance reflects real-world agent usage and informs our experimental design.

\subsection{Feature Engineering}

\subsubsection{\textbf{Feature Extraction}}
We extract 53 features spanning five categories (Table~\ref{tab:features}): commit message patterns, PR structure, code changes, patch-level code characteristics, and temporal patterns.

\begin{table}[ht]
    \centering
    \renewcommand{\arraystretch}{0}
    \vspace{-3pt}
    \caption{Feature categories extracted from PRs.}
    \vspace{-9pt}
    \label{tab:features}
    \small
    \begin{tabular}{p{2.29cm}p{5.4cm}}
    \toprule
    \textbf{Category (Count)} & \textbf{Features} \\
    \midrule
    Commit patterns (9) & Commit count, conventional commit ratio~\cite{conventional_commits}, message length (avg/min/max/std), multiline ratio, capitalization ratio \\
    \addlinespace
    PR structure (9) & Title/body length and word count, checklists, code blocks, links, bullets, conventional syntax in titles \\
    \addlinespace
    Code changes (16) & Files changed, file extensions and diversity, test/config/doc ratios, directory depth, operations (add/modify/remove/rename), additions/deletions, change concentration (Gini) \\
    \addlinespace
    Patch-level code (15) & Line counts (added/removed), line length stats, trailing whitespace, indentation, comment/import density, functions/classes/conditionals/loops \\
    \addlinespace
    Temporal (4) & Submission hour, weekend/business hours indicators, day of week\\
    \bottomrule
    \end{tabular}
    \vspace{-8pt}
\end{table}

\subsubsection{\textbf{Feature Reduction}}
To reduce multicollinearity and ensure statistical validity, we apply a two-step feature reduction:

\smallskip
\noindent\textbf{Step 1: Hierarchical clustering.}
Following standard correlation-based feature selection practices~\cite{guyon2003introduction,kuhn2013applied}, we computed pairwise absolute feature correlations and performed average-linkage hierarchical clustering. Using a threshold of $|\rho| \geq 0.70$~\cite{dormann2013collinearity}, we identified 12 highly correlated feature pairs and retained one representative from each, removing 12 redundant features. For example, \textit{body word count} is highly correlated with \textit{body length} ($\rho = 0.93$).

\smallskip
\noindent\textbf{Step 2: $R^2$ redundancy analysis.}
We identify features predictable from others using linear regression~\cite{rao2003regression}. A feature with $R^2$>0.90 when predicted from all other features is deemed redundant. After Step 1, no feature exceeded this threshold (max $R^2$=0.71), confirming that remaining features capture independent behavioral signals.

\medskip
\noindent Our final feature set includes 41 features with sufficient sample support across all classes. Following Peduzzi et al.~\cite{peduzzi1996simulation}, an events-per-variable (EPV) ratio above 10 is considered adequate for reliable model estimation. Our smallest class (Claude Code, 458 samples) achieves an EPV of 11.2, thus is less likely to suffer from overfitting.

\vspace{2pt}
\subsection{Classification Methodology}
We employ two complementary classification approaches, each using 5-fold stratified cross-validation to ensure balanced class representation across folds, and performance is reported only on held-out test folds. A preliminary experiment with SMOTE~\cite{chawla2002smote} showed it distorted the natural class distribution. We therefore report results without synthetic oversampling, reflecting the conditions practitioners would encounter in real-world deployment.

\subsubsection{\textbf{Multi-class classification}}
We train tree-based ensemble models, specifically XGBoost~\cite{chen2016xgboost} and Random Forest~\cite{breiman2001random}, to jointly predict all five agent classes, leveraging their strong performance on structured, unbalanced data~\cite{caruana2006empirical,brown2012experimental}. This approach directly assesses overall agent identification performance.

\subsubsection{\textbf{One-vs-rest binary classification}}
We train a one-vs-rest binary classifier~\cite{galar2011overview} for each agent to derive agent-specific feature importance rankings. While the multi-class model measures overall performance, this approach identifies features that distinguish individual agents, revealing their unique behavioral signatures.

\subsubsection{\textbf{Hyperparameters}}
We use moderate tree depths (max depth of 6 for XGBoost and 10 for Random Forest) and 100 estimators as baseline settings. These moderate configurations are standard in practice for balancing model expressiveness and overfitting in tree ensembles, providing shallow yet capable models that generalize well on structured data without excessive tuning.

\subsection{RQ1: How accurately can we identify which AI coding agent submitted a pull request?}

\subsubsection{\textbf{Motivation.}}

The core question for stakeholders is whether agent identification is \textit{feasible}. If agents leave no distinguishable patterns, policy enforcement and dataset validation become impossible. Conversely, high accuracy demonstrates that behavioral fingerprinting can detect undisclosed agent usage, even when developers submit agent-generated code under their own accounts. This question establishes the foundation for practical applications of our findings.

\subsubsection{\textbf{Approach.}}
We train multi-class XGBoost and Random Forest classifiers on our 41-feature set, using 5-fold stratified cross-validation. Models predict one of five agent labels for each PR. We report macro-averaged precision, recall, and F1-score across all agents, as well as per-class metrics to reveal performance variations across majority and minority classes. We also examine confusion matrices to identify systematic misclassification patterns.

\subsubsection{\textbf{Findings.}}
XGBoost outperforms Random Forest by 2.3\% F1-score (97.2\% vs 94.9\%), likely due to better handling of class imbalance through gradient boosting. Given this superior performance, we report XGBoost results for all subsequent analyses (confusion matrices, per-class metrics, feature importance).

Table~\ref{tab:multiclass} reports per-agent performance. XGBoost achieves near-perfect classification for majority classes: OpenAI Codex (99\% precision/recall), Copilot (99\%/98\%), and Devin (93\%/96\%). Cursor, a minority class, still performs well (88\%/83\%). Claude Code, the smallest class (458 samples, 1.4\% of the dataset), has lower recall (57\%) but high precision (82\%), meaning predictions of Claude Code are usually correct, but many Claude Code PRs are misclassified as other agents (mainly Devin and OpenAI Codex).

\begin{table}[ht]
    \centering
    \vspace{-5pt}
    \caption{Multi-class agent identification performance (XGBoost, 5-fold CV).}
    \vspace{-9pt}
    \label{tab:multiclass}
    \small
    \begin{tabular}{lrrrrr}
    \toprule
    \textbf{Agent} & \textbf{Samples} & \textbf{Precision} & \textbf{Recall} & \textbf{F1} & \textbf{EPV} \\
    \midrule
    OpenAI Codex   & 21,793 & 0.99 & 0.99 & 0.99 & 531.5 \\
    Copilot        &  4,967 & 0.99 & 0.98 & 0.99 & 121.1 \\
    Devin          &  4,822 & 0.93 & 0.96 & 0.94 & 117.6 \\
    Cursor         &  1,540 & 0.88 & 0.83 & 0.85 &  37.6 \\
    Claude Code    &    458 & 0.82 & 0.57 & 0.67 &  11.2 \\
    \midrule
    \textbf{Weighted Avg.} & 33,580 & \textbf{0.97} & \textbf{0.97} & \textbf{0.97} & --- \\
    \bottomrule
    \end{tabular}
    \vspace{-5pt}
\end{table}

\begin{figure}[ht]
    \centering
    \vspace{-5pt}
    \includegraphics[width=0.94\columnwidth]{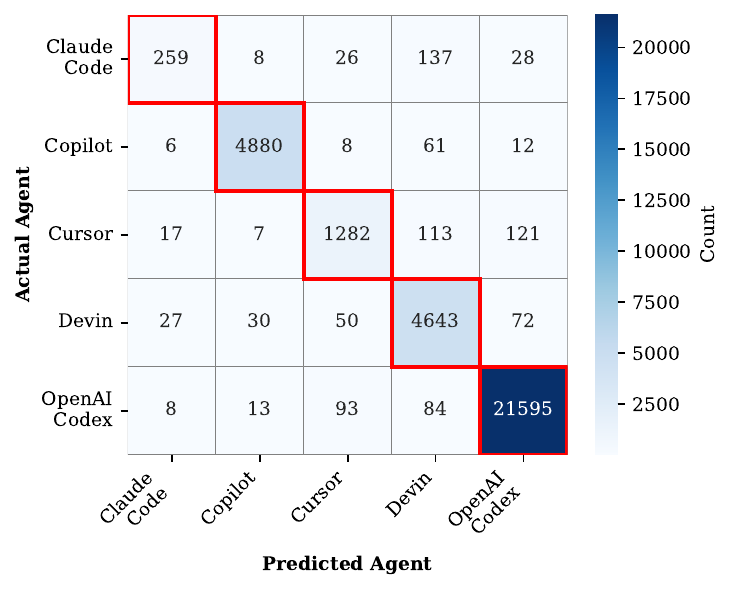}
    \vspace{-10pt}
    \caption{Confusion matrix for XGBoost multi-class agent classification. Red boxes mark correct predictions.}
    \Description{Confusion matrix showing predicted versus true agent classes for an XGBoost multi-class classifier. Rows represent actual agents and columns represent predicted agents. Cell darkness indicates the number of instances, and red boxes highlight correct predictions.}
    \label{fig:confusion}
    \vspace{-10pt}
\end{figure}

Figure~\ref{fig:confusion} shows the XGBoost multi-class confusion matrix. The strong diagonal indicates good performance, with notable confusion between Claude Code and Devin (137 cases). Claude Code PRs are mainly misclassified as Devin (137) and OpenAI Codex (28), suggesting shared characteristics despite distinct fingerprints. Cursor is mainly confused with OpenAI Codex (121) and Devin (113), indicating overlapping commit message patterns. Majority classes show little confusion: only 0.4\% of OpenAI Codex PRs are misclassified.

\smallskip
\noindent\textbf{Feature reduction robustness.} Our 53→41 feature reduction causes only 0.2\% F1-score degradation (97.4\%→97.2\%), indicating that removed features were redundant rather than discriminative. This validates our feature engineering methodology and confirms that high accuracy is not an artifact of feature multicollinearity.

\smallskip
\noindent\textbf{Model stability.} Across all five folds, XGBoost F1-score varies by only ±0.1\%, indicating consistent performance independent of train/test split. This low variance suggests our features capture stable behavioral patterns rather than dataset-specific artifacts.

\subsection{RQ2: What characteristics distinguish AI coding agents from each other?}

\subsubsection{\textbf{Motivation.}}
While RQ1 shows that agents are distinguishable, understanding how they differ is crucial so that maintainers have interpretable signals to inspect suspicious PRs, AI developers can grasp their agents’ behavioral signatures to improve authenticity, and researchers know which features enable detection to design robust comparative studies and track agent evolution.

\subsubsection{\textbf{Approach.}}

We extract feature importance from two sources:

\smallskip
\noindent\textbf{Global importance (multi-class model):} XGBoost's gain-based feature importance from the 5-class classifier reveals which features best discriminate any agent from others. This identifies universal patterns across all agents.

\smallskip
\noindent\textbf{Agent-specific importance (one-vs-rest models):} For each agent, we train a binary XGBoost classifier (agent $x$ versus all others) and extract feature importance. This reveals unique signatures for each AI coding agent, especially important for minority classes whose patterns may be dominated by majority classes in multi-class importance rankings.
We report the top three features for each agent from one-vs-rest models, providing actionable fingerprints for practitioners.

\medskip
\subsubsection{\textbf{Findings.}}
~

\smallskip
\noindent\textbf{Global patterns.}
The global feature importance from our multi-class XGBoost model revealed that \emph{Commit} message characteristics dominate: \emph{multiline commit ratio} (44.7\%), followed by \emph{change concentration gini} (10.1\%) and \emph{average commit message length} (2.3\%). Notably, code content features (comment density, conditionals, functions) rank lower, indicating that how agents communicate changes is more distinctive than what changes they make. This finding has practical implications: agents may generate similar code but differ in version control and communication patterns.

\begin{figure}[ht]
    \centering
    \vspace{-2pt}
    \includegraphics[width=\linewidth]{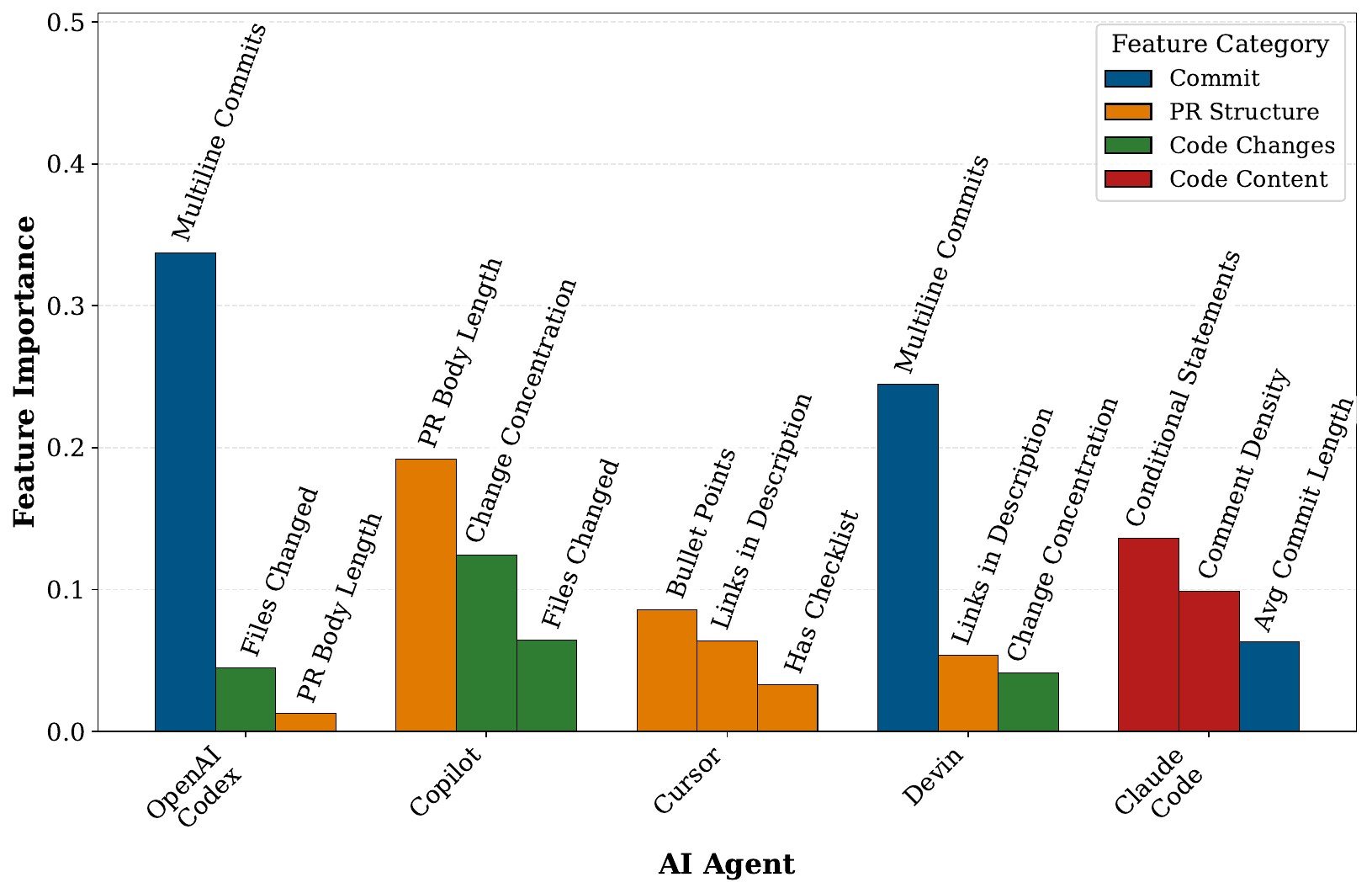}
    \vspace{-20pt}
    \caption{Agent-specific feature importance from one-vs-rest XGBoost models.}
    \Description{Bar charts showing agent-specific feature importance scores derived from one-vs-rest XGBoost models. Each subplot corresponds to one agent, with bars representing the relative contribution of individual features to the classifier.}
    \label{fig:signatures}
    \vspace{-10pt}
\end{figure}

\smallskip
\noindent\textbf{Agent-specific signatures.} 
Figure~\ref{fig:signatures} shows the top 3 discriminative features for each agent from one-vs-rest XGBoost models. We observe that OpenAI Codex is distinguished by extensive multiline commit messages (67.5\%), producing more detailed commit text than other agents, likely reflecting an emphasis on comprehensive documentation. Copilot is associated with longer PR descriptions (38.4\%) and higher change concentration (24.9\%), indicating focused code changes with detailed explanations. Cursor frequently uses bullet points (17.2\%) and hyperlinks (12.8\%) in PR bodies, yielding highly structured, reference-rich descriptions, likely shaped by its interface. Devin combines multiline commit messages (48.9\%) with more distributed changes across files (8.2\%), suggesting an autonomous workflow with granular commits. Claude Code shows distinctive code-level traits, including high conditional density (27.2\%) and elevated comment density (19.8\%), consistent with well-documented, control-flow-intensive code.

\smallskip
\noindent\textbf{One-vs-rest reveals minority class patterns.}
In multi-class importance, Claude Code's distinctive features (conditionals, comments) rank 15th and 24th globally, dominated by majority class patterns. This occurs because multi-class feature importance is biased toward the largest class (OpenAI Codex, 64.9\% of the dataset), which inflates features that distinguish it at the expense of minority classes. One-vs-rest analysis elevates these features to top importance, demonstrating that minority classes require targeted analysis to reveal their unique fingerprints.

\section{Implications}
\label{sec:implications}
\vspace{-1pt}
\noindent\textbf{For Project Maintainers.}  
Our models, with a 97.2\% F1-score, flag likely AI-generated PRs for review, enabling enforceable AI usage policies. Agent-specific signatures (e.g., Cursor’s bullet-heavy descriptions, Claude Code’s dense comments) offer interpretable cues for non-experts. Automated pre-merge checks can make such policies verifiable. Maintainers should note that adversaries may mimic human patterns, requiring continual model updates.

\smallskip
\noindent\textbf{For AI Agent Developers.}  
Our findings reveal that AI coding agents have distinct behavioral signatures that set them apart. Developers can use these insights to refine agent design: those prioritizing transparency can preserve or highlight unique patterns in commit messages, PR templates, and code formatting to clearly signal AI authorship. Conversely, agents that want to blend with normal workflows can vary their commit structure, PR format, and code style to be less distinctive. Recognizing these patterns enables deliberate choices about agent behavior and disclosure.

\smallskip
\noindent\textbf{For Researchers.}  
Our findings challenge the validity of datasets separating ``\textit{human}'' and AI contributions. For example, AIDev~\cite{li2025aidev} labels 6,618 PRs as ``\textit{human}'', yet developers may have used AI coding agents, thus introducing noise that can mislead empirical analyses~\cite{ghaleb2019noise} on code quality, bug density, or development velocity. Researchers should (1) apply fingerprinting to validate labels, (2) note contamination as a limitation, or (3) use pre-agent datasets. Longitudinal studies should track whether fingerprints remain stable as agents and developers adapt.

\smallskip
\noindent\textbf{Overall}, our work raises key methodological questions: How stable are these fingerprints across languages, domains, and repositories? Do agents leave different signatures in greenfield versus maintenance work? Can combining behavioral signals with code-level LLM detection~\cite{tian2023chatgpt} improve accuracy? These questions frame a research agenda on AI’s impact on software development.

\vspace{-2pt}
\section{Threats to Validity}
\label{sec:threats}
\vspace{-1pt}

\noindent\textbf{Construct Validity.}
Our features assume agents behave consistently, but agent developers may intentionally or unintentionally modify these patterns over time. We cannot detect such evolution in our cross-sectional dataset. Future work should examine temporal stability of fingerprints using time-split validation (train on early PRs, test on later ones) to measure signature drift.

\smallskip
\noindent\textbf{Internal Validity.}
Class imbalance affects minority class performance. Claude Code (458 samples, EPV=11.2) achieves only 57\% recall despite meeting statistical thresholds~\cite{peduzzi1996simulation}. We avoided SMOTE to preserve real-world conditions, but practitioners needing high recall for specific agents could benefit from oversampling. In addition, we use Random Forest and XGBoost with conservative depth limits to avoid deep tree memorization, yet other models (e.g., neural networks) may yield different performance-interpretability trade-offs.

\smallskip
\noindent\textbf{External Validity.}
Our dataset consists of public GitHub repositories where AI coding agents submit PRs directly. We do not test generalizability to private repositories or other platforms (e.g., GitLab), where agent behavior may differ due to stricter style guides or organizational practices. Adversaries aware of our fingerprinting methods could attempt to evade detection, but this would require sophisticated knowledge of feature importance and likely represents a small fraction of undisclosed agent usage. Also, as AI coding agents evolve, fingerprints may change, thus requiring model retraining.

\section{Related Work}
\label{sec:related}

\vspace{-1pt}
\noindent\textbf{AI-Generated Code Detection.} 
Prior work detects AI-generated code using LLM-based approaches~\cite{tian2023chatgpt,Nguyen2024gptsniffer} or stylometric analysis~\cite{bisztray2025know}. Tian et al.~\cite{tian2023chatgpt} achieve 93\% accuracy distinguishing ChatGPT-generated from human code using perplexity and burstiness metrics. These approaches focus on separating AI from human code, not identifying \textit{which} agent generated it. Our work advances this line by fingerprinting individual agents, enabling granular policy enforcement and dataset validation.

\smallskip
\noindent\textbf{Code Authorship Attribution.} 
Code authorship attribution uses features like identifier naming, code structure, and documentation to identify developers~\cite{caliskan2015anonymizing,burrows2007source}. Recent work employed deep learning~\cite{abuhamad2018large} and graph neural networks~\cite{zhang2019novel}. We adapt these techniques to AI coding agents and find that commit message conventions are more discriminative than code changes, unlike in human authorship studies where code style dominates~\cite{caliskan2015anonymizing}, indicating agents exhibit distinct version-control behaviors beyond code generation.

\smallskip
\noindent\textbf{Bot Detection in Software Engineering.} 
Prior work identifies automated accounts (CI bots, dependency bots) using metadata and commit patterns~\cite{wessel2018power}. Golzadeh et al.~\cite{golzadeh2022accuracy}  detect bot accounts via message templates and timing. Unlike these intentionally identifiable bots, our model can detect AI usage when developers submit PRs under their own accounts, relying on behavioral patterns rather than metadata. Even without account-level signals, commit and PR features reveal agent involvement.

\section{Conclusion}
\label{sec:conclusion}
We present the first study on AI coding agent authorship attribution through behavioral fingerprinting, analyzing 33,580 PRs from five major agents. Using 41 features capturing commit patterns, PR structure, and code characteristics, our XGBoost classifier achieves 97.2\% F1-score in identifying the submitting agent. Distinctive signatures include Codex’s multiline commits (67.5\%), Copilot’s detailed PR bodies (38.4\%), Cursor’s heavy descriptions (17.2\%), Devin’s conventional commits (48.9\%), and Claude Code’s code complexity patterns (27.2\% conditionals, 19.8\% comments).
These fingerprints enable maintainers to enforce AI usage policies, help developers make agent outputs more transparent, and alert researchers to contamination in supposedly “human” datasets, such as AIDev’s 6,618 PRs.
Our results show that AI coding agents leave strong behavioral fingerprints, enabling the detection of both autonomous agent PRs and human-submitted PRs with agent-generated code, thus improving repository governance, research validity, and analysis of AI’s impact on software development.

\medskip
\noindent\textbf{Future work} includes studying the temporal stability of fingerprints, expanding to more agents and languages, improving robustness to adversarial behavior, combining behavioral and code-level detection to enhance accuracy, and validating on human-submitted PRs containing agent-generated code to test provenance claims.

\vspace{-1pt}
\begin{acks}
\vspace{-1pt}
This work is funded by the Natural Sciences and Engineering Research Council of Canada (NSERC): RGPIN-2025-05897.
\end{acks}
\vspace{-1pt}

\clearpage
\balance
\bibliographystyle{ACM-Reference-Format}
\bibliography{refs}

@article{peduzzi1996simulation,
  title={A simulation study of the number of events per variable in logistic regression analysis},
  author={Peduzzi, Peter and Concato, John and Kemper, Elizabeth and Holford, Theodore R and Feinstein, Alvan R},
  journal={Journal of clinical epidemiology},
  volume={49},
  number={12},
  pages={1373--1379},
  year={1996},
  publisher={Elsevier}
}

@article{li2025aidev,
  title={{The Rise of AI Teammates in Software Engineering (SE) 3.0: How Autonomous Coding Agents Are Reshaping Software Engineering}}, 
  author={Li, Hao and Zhang, Haoxiang and Hassan, Ahmed E.},
  journal={arXiv preprint arXiv:2507.15003},
  year={2025}
}

@misc{github_tos,
  title={GitHub Terms of Service},
  author={{GitHub}},
  year={2024},
  howpublished={\url{https://docs.github.com/en/site-policy/github-terms/github-terms-of-service}}
}

@article{barke2023grounded,
  title={Grounded {Copilot}: How programmers interact with code-generating models},
  author={Barke, Shraddha and James, Michael B and Polikarpova, Nadia},
  journal={Proceedings of the ACM on Programming Languages},
  volume={7},
  number={OOPSLA1},
  pages={85--111},
  year={2023},
  publisher={ACM New York, NY, USA}
}

@misc{conventional_commits,
  title={Conventional Commits Specification},
  author={{Conventional Commits}},
  year={2024},
  howpublished={\url{https://www.conventionalcommits.org/}}
}

@misc{rao2003regression,
  title={Regression modeling strategies: with applications to linear models, logistic regression, and survival analysis},
  author={Rao, Sunil J},
  year={2003},
  publisher={Taylor \& Francis}
}

@inproceedings{tian2023chatgpt,
  title={Is {ChatGPT} the Ultimate Programming Assistant -- How far is it?},
  author={Haoye Tian and Weiqi Lu and Tsz On Li and Xunzhu Tang and Shing-Chi Cheung and Jacques Klein and Tegawendé F. Bissyandé},
  booktitle={arXiv preprint arXiv:2304.11938},
  year={2023}
}

@article{Nguyen2024gptsniffer,
  title={{GPTSniffer}: A {CodeBERT-based} classifier to detect source code written by {ChatGPT}},
  author={Nguyen, Phuong T and Di Rocco, Juri and Di Sipio, Claudio and Rubei, Riccardo and Di Ruscio, Davide and Di Penta, Massimiliano},
  journal={Journal of Systems and Software},
  volume={214},
  pages={112059},
  year={2024},
  publisher={Elsevier}
}

@inproceedings{bisztray2025know,
  title={I Know Which LLM Wrote Your Code Last Summer: LLM generated Code Stylometry for Authorship Attribution},
  author={Bisztray, Tamas and Cherif, Bilel and Dubniczky, Richard A and Gruschka, Nils and Borsos, Bertalan and Ferrag, Mohamed Amine and Kovacs, Attila and Mavroeidis, Vasileios and Tihanyi, Norbert},
  booktitle={Proceedings of the 18th ACM Workshop on Artificial Intelligence and Security},
  pages={28--39},
  year={2025}
}

@inproceedings{caliskan2015anonymizing,
  title={De-anonymizing programmers via code stylometry},
  author={Caliskan-Islam, Aylin and Harang, Richard and Liu, Andrew and Narayanan, Arvind and Voss, Clare and Yamaguchi, Fabian and Greenstadt, Rachel},
  booktitle={24th USENIX security symposium (USENIX Security 15)},
  pages={255--270},
  year={2015}
}

@inproceedings{burrows2007source,
  title={Source code authorship attribution using n-grams},
  author={Burrows, Steven and Tahaghoghi, Seyed MM},
  booktitle={Proceedings of the twelth Australasian document computing symposium, Melbourne, Australia, RMIT University},
  pages={32--39},
  year={2007},
  organization={Citeseer}
}

@inproceedings{abuhamad2018large,
  title={Large-scale and language-oblivious code authorship identification},
  author={Abuhamad, Mohammed and AbuHmed, Tamer and Mohaisen, Aziz and Nyang, DaeHun},
  booktitle={Proceedings of the 2018 ACM SIGSAC Conference on Computer and Communications Security},
  pages={101--114},
  year={2018}
}

@inproceedings{zhang2019novel,
  title={A novel neural source code representation based on abstract syntax tree},
  author={Zhang, Jian and Wang, Xu and Zhang, Hongyu and Sun, Hailong and Wang, Kaixuan and Liu, Xudong},
  booktitle={2019 IEEE/ACM 41st International Conference on Software Engineering (ICSE)},
  pages={783--794},
  year={2019},
  organization={IEEE}
}

@article{wessel2018power,
  title={The power of bots: Characterizing and understanding bots in {OSS} projects},
  author={Wessel, Mairieli and De Souza, Bruno Mendes and Steinmacher, Igor and Wiese, Igor S and Polato, Ivanilton and Chaves, Ana Paula and Gerosa, Marco A},
  journal={Proceedings of the ACM on Human-Computer Interaction},
  volume={2},
  number={CSCW},
  pages={1--19},
  year={2018},
  publisher={ACM New York, NY, USA}
}

@inproceedings{golzadeh2022accuracy,
  title={On the accuracy of bot detection techniques},
  author={Golzadeh, Mehdi and Decan, Alexandre and Chidambaram, Natarajan},
  booktitle={Proceedings of the Fourth International Workshop on Bots in Software Engineering},
  pages={1--5},
  year={2022}
}

@misc{replication_package,
  title={Fingerprinting AI Coding Agents on GitHub (Replication Package)},
  author={Taher A. Ghaleb},
  howpublished={\url{https://github.com/Taher-Ghaleb/AIAgentsFingerprinting-MSR2026}},
  year={2026}
}

@article{guyon2003introduction,
  title={An introduction to variable and feature selection},
  author={Guyon, Isabelle and Elisseeff, Andr{\'e}},
  journal={Journal of machine learning research},
  volume={3},
  number={Mar},
  pages={1157--1182},
  year={2003}
}

@book{kuhn2013applied,
  title={Applied predictive modeling},
  author={Max Kuhn and Kjell Johnson},
  year={2013},
  publisher={Springer}
}

@article{breiman2001random,
  title={Random forests},
  author={Breiman, Leo},
  journal={Machine learning},
  volume={45},
  number={1},
  pages={5--32},
  year={2001},
  publisher={Springer}
}

@article{chen2016xgboost,
  title={XGBoost: A Scalable Tree Boosting System},
  author={Chen, Tianqi},
  journal={Cornell University},
  year={2016}
}

@inproceedings{caruana2006empirical,
  title={An empirical comparison of supervised learning algorithms},
  author={Caruana, Rich and Niculescu-Mizil, Alexandru},
  booktitle={Proceedings of the 23rd international conference on Machine learning},
  pages={161--168},
  year={2006}
}

@article{brown2012experimental,
  title={An experimental comparison of classification algorithms for imbalanced credit scoring data sets},
  author={Brown, Iain and Mues, Christophe},
  journal={Expert systems with applications},
  volume={39},
  number={3},
  pages={3446--3453},
  year={2012},
  publisher={Elsevier}
}

@article{dormann2013collinearity,
  title={Collinearity: a review of methods to deal with it and a simulation study evaluating their performance},
  author={Dormann, Carsten F and Elith, Jane and Bacher, Sven and Buchmann, Carsten and Carl, Gudrun and Carr{\'e}, Gabriel and Marqu{\'e}z, Jaime R Garc{\'\i}a and Gruber, Bernd and Lafourcade, Bruno and Leit{\~a}o, Pedro J and others},
  journal={Ecography},
  volume={36},
  number={1},
  pages={27--46},
  year={2013},
  publisher={Wiley Online Library}
}

@article{chawla2002smote,
  title={{SMOTE:} synthetic minority over-sampling technique},
  author={Chawla, Nitesh V and Bowyer, Kevin W and Hall, Lawrence O and Kegelmeyer, W Philip},
  journal={Journal of artificial intelligence research},
  volume={16},
  pages={321--357},
  year={2002}
}

@article{galar2011overview,
  title={An overview of ensemble methods for binary classifiers in multi-class problems: Experimental study on one-vs-one and one-vs-all schemes},
  author={Galar, Mikel and Fern{\'a}ndez, Alberto and Barrenechea, Edurne and Bustince, Humberto and Herrera, Francisco},
  journal={Pattern Recognition},
  volume={44},
  number={8},
  pages={1761--1776},
  year={2011},
  publisher={Elsevier}
}

@article{ghaleb2019noise,
  title={Studying the impact of noises in build breakage data},
  author={Ghaleb, Taher Ahmed and Da Costa, Daniel Alencar and Zou, Ying and Hassan, Ahmed E},
  journal={IEEE Transactions on Software Engineering},
  volume={47},
  number={9},
  pages={1998--2011},
  year={2019},
  publisher={IEEE}
}

\end{document}